\documentclass{aastex62}
\usepackage{amsmath}
\usepackage{epstopdf}
\usepackage{float}
\usepackage{graphicx}
\usepackage{listings}
\usepackage{rotating}

\newcommand{\lapprox} {\, \lower3pt\hbox{$\sim$}\llap{\raise2pt\hbox{$<$}}\,}
\newcommand{\gapprox} {\, \lower3pt\hbox{$\sim$}\llap{\raise2pt\hbox{$>$}}\,}

\begin{document}

\title{Scaling Laws for Dynamic Solar Loops}

\author[0000-0002-3300-6041]{Stephen J. Bradshaw}
\affiliation{Department of Physics \& Astronomy, Rice University, Houston, TX 77005, USA}

\author[0000-0001-8720-0723]{A. Gordon Emslie}
\affiliation{Department of Physics \& Astronomy, Western Kentucky University, Bowling Green, KY 42101, USA}

\begin{abstract}
The scaling laws which relate the peak temperature $T_M$ and volumetric heating rate $E_H$ to the pressure $P$ and length $L$ for static coronal loops were established over 40 years ago; they have proved to be of immense value in a wide range of studies. Here we extend these scaling laws to {\it dynamic} loops, where enthalpy flux becomes important to the energy balance, and study impulsive heating/filling characterized by upward enthalpy flows. We show that for collision-dominated thermal conduction, the functional dependencies of the scaling laws are the same as for the static case, when the radiative losses scale as $T^{-1/2}$, but with a different constant of proportionality that depends on the Mach number $M$ of the flow. The dependence on the Mach number is such that the scaling laws for low to moderate Mach number flows are almost indistinguishable from the static case. When thermal conduction is limited by turbulent processes, however, the much weaker dependence of the scattering mean free path (and hence thermal conduction coefficient) on temperature leads to a limiting Mach number for return enthalpy fluxes driven by thermal conduction between the corona and chromosphere.
\end{abstract}

\keywords{Sun: corona -- Sun: coronal loops}

\section{Introduction}\label{introduction}

The solar corona is structured by magnetic fields which confine plasma at multi-million degree temperatures. Individual loop-like structures appear in contrast against the background corona when they are heated and filled with high temperature and density plasma. The mechanism of coronal heating remains a largely open question, although there exists a broad consensus that it must involve the conversion and dissipation of excess magnetic energy. The energy balance in the corona chiefly involves an interplay between energy input by the aforementioned unidentified mechanism, energy loss by optically-thin radiation, and the internal redistribution of energy by thermal conduction and bulk flows driven by excess pressures. \cite{1978ApJ...220..643R} used Skylab X-ray observations of coronal loops, assuming the structures were both quasi-static (neglecting bulk flows) and isobaric, and the energy balance between heating, radiation, and thermal conduction, to find their eponymous scaling laws which connect a coronal loop's maximum temperature $T_M$ (K) and uniform volumetric heating rate $E_H$ (erg~cm$^{-3}$~s$^{-1}$) to its pressure $P$ (dyne~cm$^{-2}$) and half-length $L$ (cm).

\begin{equation}\label{rtv43}
T_M \simeq 1.4 \times 10^3 \, (PL)^{1/3}
\end{equation}
(their Equation~(4.3)) and

\begin{equation}\label{rtv44}
E_H \simeq 9.8 \times 10^4 \, P^{7/6} \, L^{-5/6}
\end{equation}
(their Equation~(4.4)). Their modeling assumed heat conduction proportional to the temperature gradient and dominated by collisional transport predominantly between electrons and, we reiterate, a static environment in which energy transport through mass motions was not considered. See also \cite{1978A&A....70....1C} and \cite{2010ApJ...714.1290M} while reviews are given by \cite{2004psci.book.....A} and \cite{2014LRSP...11....4R}.

Heating in the solar corona is expected to quickly raise the temperature of the plasma, due to its tenuous nature and low heat capacity, and drive strong thermal conduction fronts away from the heating site, towards the lower atmosphere, in an effort to efficiently shed the excess energy that cannot be radiated away. As the conduction fronts reach the transition region and the plasma density increases, radiation begins to remove the incoming heat flux and thermal conduction becomes an increasingly less efficient energy transport mechanism.  This leads to energy deposition into the plasma and a localized increase in temperature, which steepens the temperature gradient. The increase in temperature also increases the local pressure, driving a flow of heated material back into the corona and filling the loop. This return enthalpy flux comprises a major contribution to the coronal energy balance.

\cite{2013ApJ...770...12B} undertook a detailed study of this process and demonstrated the crucial need to resolve the steep transition region gradients in order to properly capture the complicated interplay of energy and momentum transport in numerical models.  However, significant physical insight can nevertheless be derived by considering the scaling of dominant terms in the energy equation with physical parameters such as temperature and pressure.  This is perhaps best exemplified by ``zero-dimensional'' (0D) loop models, in particular the ``enthalpy-based thermal evolution of loops'' (EBTEL) model of \citet{2008ApJ...682.1351K} and \citet{2012ApJ...758....5C}, that establish the behavior of loop-averaged quantities as functions of time. Given the significantly different importance of the various terms in the energy equation when strong mass motions are present, it is therefore of interest to examine the form of the scaling laws for {\it dynamic} loops.

In Section~\ref{analysis} we first discuss the range of validity for the dynamic scaling laws we subsequently derive, showing that an upper limit to the Mach number exists for a hydrodynamic flow under the conditions we are considering, which confines it to the subsonic regime, and present the form of the energy equations and the resulting 0D dynamic scaling laws. We find that the functional dependencies of the maximum loop temperature and uniform volumetric heating rate are, quite remarkably, {\it the same as for the static case}, but only when the radiative losses scale as $T^{-1/2}$. Further, the multiplicative coefficient in the scaling laws (Equations~(\ref{rtv43}) and (\ref{rtv44})) increases with the Mach number for subsonic flows.

Because it is possible that the energy release associated with coronal heating creates significant turbulence in the surrounding medium, we find dynamic scaling laws associated with turbulence-dominated thermal conduction \citep[extending][]{2019ApJ...880...80B} in Section~\ref{turb-dom}. In such regimes, consideration of the ways in which the heat flux could drive increasingly fast upward enthalpy fluxes leads to the conclusion that flow velocity is in fact constrained by a limit on the Mach number, which can be smaller than the limit imposed by hydrodynamics alone.

In Section~\ref{conclusions}, we summarize the results obtained and discuss their significance for the study of dynamic loop structures.

\section{Derivation of the Scaling Laws for Dynamic Solar Loops}
\label{analysis}

We begin in Section~\ref{validity} by establishing the conditions under which the dynamic scaling laws we will derive are valid. This is necessary because, unlike the static treatment followed by \cite{1978ApJ...220..643R} \citep[and][]{2019ApJ...880...80B}, one cannot assume uniform pressure along the loop, particularly as the flow speed increases and its kinetic energy approaches the plasma thermal energy. Furthermore, in the dynamic case, it is necessary to consider the timescale on which a steady flow can be established.

In Sections~\ref{scaling_law_1} and \ref{scaling_law_2}, beginning with the hydrodynamic energy equation for a coronal loop which describes the balance between energy in, radiation out, and internal redistribution by thermal conduction and mass motions \citep[the term omitted by ][]{1978ApJ...220..643R}, we derive the dynamic scaling laws for the maximum (apex) temperature $T_M$ and the uniform volumetric heating rate $E_H$. In this section we consider the limit in which thermal conduction is dominated by Coulomb collisions and in Section~\ref{turb-dom} we consider the case of thermal conduction strongly suppressed by turbulence \citep[see, e.g.,][]{2018ApJ...852..127B}, generalizing the treatment of \cite{2019ApJ...880...80B} to dynamic loops.

\subsection{The Validity of Dynamic Scaling Laws}
\label{validity}

We consider a field-aligned geometry in which each quantity is expressed as a function of a spatial coordinate $s$ (cm), along a guiding magnetic field, in the direction of the coronal loop apex. In the presence of flows, particularly strong flows, a Bernoulli condition

\begin{equation}\label{}
P + P_D = {\rm const.}
\end{equation}
must be satisfied along each streamline.  Here $P = 2k_B n T$ (dyne~cm$^{-2}$) is the static pressure (associated with the state of the medium rather than its movement) and $P_D = (1/2) n m_p V^2$ is the dynamic pressure, $k_B = 1.38 \times 10^{-16}$~erg~K$^{-1}$ is Boltzmann's constant, $m_p$ (g) is the proton mass, and $n$ (cm$^{-3}$) and $T$ (K) are the electron density and temperature, respectively. In differential form, the flow must satisfy the hydrodynamic equations which describe the conservation of mass

\begin{equation}\label{mass-cons}
\frac{\partial}{\partial s} \left( \rho V \right) = 0 \,\,\, ,
\end{equation}
and momentum

\begin{equation}\label{mom-cons}
\frac{\partial}{\partial s} \left( \rho V^2 \right) = -\frac{\partial P}{\partial s} \,\,\,
\end{equation}
in steady-state ($\partial/\partial t = 0$). It is convenient to eliminate the bulk flow velocity $V$ (cm~s$^{-1}$) as a variable by writing it in terms of the Mach number $M$ and the sound speed $C_S = \sqrt{\gamma P / \rho}$. Substituting $V = MC_S$ and the mass density $\rho = m_p n$ (g~cm$^{-3}$) into Equations~(\ref{mass-cons}) and~(\ref{mom-cons}), we find

\begin{equation}\label{mass-cons2}
\frac{\partial}{\partial s} \left( M n T^{1/2} \right) = 0
\end{equation}
and

\begin{equation}\label{mom-cons2}
\frac{\partial}{\partial s} \left( [\gamma M^2 +1 ] \, n T \right) = 0 \,\,\, .
\end{equation}
Note that we have eliminated $P$ in favor of working with $n$ from here on. While the scaling laws have traditionally been formulated with respect to $P$, its non-uniform nature in the dynamic case offers no convenient advantage over $n$, and the latter quantity is more directly accessible and thus commonly used in observational contexts \citep[e.g.,][]{2018LRSP...15....5D}. Nonetheless, we provide formulations of all scaling laws with respect to $P$ in Appendices~\ref{app_a} and \ref{app_b}, with the proviso that it must no longer be considered a constant but, rather, the value at the maximum (apex) temperature (i.e., $P_M$).

Expanding Equations~(\ref{mass-cons2}) and (\ref{mom-cons2}) gives

\begin{equation}\label{mass-cons3}
\partial M = - M \left( \frac{1}{2} \, \partial \ln T + \partial \ln n \right)
\end{equation}
and

\begin{equation}\label{mom-cons3}
\frac{2 \gamma M}{\gamma M^2 +1} \, \partial M = - ( \partial \ln T + \partial \ln n ) \,\,\, .
\end{equation}
Eliminating $\partial M$ between Equations~(\ref{mass-cons3}) and~(\ref{mom-cons3}) gives the relation between temperature and density variations and the Mach number $M$:

\begin{equation}\label{nTM-eqn}
\left( \gamma M^2 - 1 \right) \partial \ln n = \partial \ln T \,\,\, .
\end{equation}

In the static case $M=0$ and we recover the constant pressure condition $nT=$~constant, which is simply equivalent to setting the left-hand side of Equation~(\ref{mom-cons}) to zero. More interestingly, since we expect $T$ to increase with altitude ($\partial \ln T /\partial s > 0$) and $n$ to decrease with altitude ($\partial \ln n/\partial s < 0$), then we require $\gamma M^2 < 1$, which places a strong constraint on $M$. With the ratio of specific heats $\gamma = 5/3$, the upper-limit on the Mach number is

\begin{equation}\label{mmax-hydro}
M_{\rm max} = \sqrt{1/\gamma} \simeq 0.77 \,\,\, .
\end{equation}
The corresponding maximum flow speed is given by $V_{\rm max} = M_{\rm max} C_S = \sqrt{1/\gamma} \sqrt{2\gamma k_B T/m_p} = \sqrt{2k_B T/m_p}$, which we recognize as the ion thermal speed in the plasma. This is consistent with the solution of \cite{1958ApJ...128..664P} (specifically, Equation~(18) in that paper) for a steady-state outflow in one dimension (specified in cartesian coordinates, rather than a 1D spherically symmetric outflow). Thus, the flow speed is limited to $V \simeq 0.77 C_S$, or simply the ion thermal speed, everywhere along the loop. Note that transonic solutions to Equation~(\ref{nTM-eqn}) also exist if $M$ passes through the value $M_{\rm max}$ at the (critical) point where the temperature gradient vanishes ($\partial \ln T = 0$) \citep[][]{1968AJS....73Q..72M,1980SoPh...65..251C,1982GApFD..20..227C,1995A&A...294..861O}. However, since we are interested in coronal loops with monotonically increasing temperature (see above), and not in the isothermal case considered by \cite{1958ApJ...128..664P}, we do not consider these solutions.

Finally, we consider the timescale required to establish a dynamic steady-state. The bulk flow is limited to roughly three-quarters of the sound speed and so the flow of information through the system is guaranteed to always be faster ($L/C_S < L/V$). Consequently, a time interval will exist during which a (quasi) steady-state in the coronal loop can be assumed. Clearly, this is a more robust assumption for weaker flows and one must be cautious when applying the dynamic scaling laws in the presence of relatively fast flows (e.g., $M \gapprox 0.5$).

\subsection{Temperature, Density, and Length}
\label{scaling_law_1}

As shown by the formal derivations of \cite{2010ApJ...714.1290M} and \cite{2019ApJ...880...80B}, the first scaling law of \cite{1978ApJ...220..643R} (Equation~(\ref{rtv43}) above) can be well approximated by setting the energy redistributed by thermal conduction equal to the radiated energy, using the same temperature $T_M$ for both processes;  In the static case the transition region is assumed to radiate all of the incoming heat flux from the corona and so the heat flux at the lower boundary is set to zero, and indeed one could define the transition region length as the distance over which the coronal heat flux is completely radiated. By contrast, in the dynamic case the incoming heat flux is partitioned between radiation and a return enthalpy flux back into the corona, comprising transition region material heated and ablated (``evaporated'') by thermal conduction \citep{2013ApJ...770...12B} which then fills the loop. The energy balance is given by

\begin{equation}\label{energy_equation}
\frac{dF_E}{ds} + E_R = -\frac{dF_C}{ds} \,\,\, .
\end{equation}
Here

\begin{equation}\label{enthalpy_flux}
F_E = \left ( \frac{\gamma}{\gamma-1} \, P + \frac{1}{2} \, \rho \, V^2 \right ) \, V \,\,\,
\end{equation}
is the enthalpy flux (erg~cm$^{-2}$~s$^{-1}$) composed of thermal and kinetic energy components. The radiative losses per unit volume (erg~cm$^{-3}$~s$^{-1}$) are given by the (optically thin) expression

\begin{equation}\label{radiative_losses}
E_R = n^2 \Lambda(T).
\end{equation}
The quantity $\Lambda(T)$ (erg~cm$^3$~s$^{-1}$) is the emissivity (or loss function) for optically-thin radiation. This is often presented as a set of piece-wise power-laws or a look-up table in numerical treatments, but in the temperature range of interest to us ($10^5$~K~$ \lapprox T \lapprox 10^7$~K) can be well approximated by a single power-law

\begin{equation}\label{rad-losses-power-law}
\Lambda(T) = \chi \, T^\alpha \,\,\, ,
\end{equation}
with parameter values $\chi \simeq 1.6 \times 10^{-19}$~erg~cm$^3$~s$^{-1}$~K$^{1/2}$ and $\alpha=-1/2$.

We set the heat flux term $F_C$ (erg~cm$^{-2}$~s$^{-1}$) to a form appropriate to the situation where the mean free path $\lambda$ (cm) pertinent to the transport process is small compared to the overall scale $L$ (cm) of the problem (the Knudsen number ${\rm Kn} \equiv \lambda/L \ll 1$).  For such a situation, the heat flux is proportional to the local temperature gradient

\begin{equation}\label{cond-general}
F_C = - \kappa (n, T) \, \frac{dT}{ds} \,\,\, ,
\end{equation}
with the coefficient $\kappa$ (erg~cm$^{-1}$~K$^{-7/2}$~s$^{-1}$) related to the mean free path $\lambda$ by

\begin{equation}\label{kappa-general}
\kappa = 2 n k_B V_{\rm th} \, \lambda \,\,\, ,
\end{equation}
where $V_{\rm th}= \sqrt{2 k_B T/m_e}$ (cm~s$^{-1}$) is the thermal speed (of electrons, in this case). For heat transport dominated by Coulomb collisions, the appropriate value for $\lambda$ is the collisional mean free path \citep[e.g.,][]{1962pfig.book.....S}:

\begin{equation}\label{lambda-collisional}
\lambda_C = \frac{(2k_B T)^2}{2\pi e^4 \ln \Lambda \, n} \simeq 10^4 \, \frac{T^2}{n} \,\,\, ,
\end{equation}
where $e = 4.8 \times 10^{-10}$~esu is the electronic charge and $\ln \Lambda \simeq 20$ is the Coulomb logarithm. Thus

\begin{equation}\label{kappa-spitzer}
\kappa = \frac{2n k_B(2k_BT)^{1/2}}{m_e^{1/2}} \, \lambda_C = \frac{k_B \, (2 k_B)^{5/2}}{\pi m_e^{1/2} e^4 \ln \Lambda} \, T^{5/2} = \kappa_0 \, T^{5/2} \,\,\, ,
\end{equation}
where $\kappa_0 \simeq 1.7 \times 10^{-6}$~erg~cm$^{-1}$~s$^{-1}$~K$^{-7/2}$ \citep{2019ApJ...880...80B}.  The heat flux can thus be written in the usual form

\begin{equation}\label{conduction-spitzer}
F_{C} = - \kappa_0 \, T^{5/2} \, \frac{dT}{ds} = - \frac{2}{7} \, \kappa_0 \, \frac{dT^{7/2}}{ds}\,\,\, .
\end{equation}

Integrating the energy balance equation~(\ref{energy_equation}) over one side of a coronal loop (half-length $L$) gives the zero-dimensional expression

\begin{equation}\label{0-D}
\left ( \frac{\gamma}{\gamma-1} \, P + \frac{1}{2} \, \rho \, M^2 C_S^2 \right ) \, M \,  C_S + n_M^2 \, \Lambda(T_M) \, L = \frac{2}{7} \, \kappa_0 \, \frac{T_M^{7/2}}{L} \,\,\, ,
\end{equation}
where we have again eliminated $V$ by writing it in terms of the Mach number and sound speed. We note that all terms in Equation~(\ref{0-D}) involve conditions at the apex of the loop, a simplification justified by the high sensitivity of the heat flux to temperature and supported by a more exact treatment \citep[e.g.,][]{2010ApJ...714.1290M,2019ApJ...880...80B}. Substituting for the mass density and the pressure casts Equation~(\ref{0-D}) in the form

\begin{equation}\label{energy-equation-red}
\left ( \frac{8 \gamma^3 k_B^3}{m_p} \right )^{1/2}  \, \left ( \frac{M}{\gamma-1} + \frac{M^3}{2} \right ) + \chi \, n_M \, T_M^{-2} \, L  = \frac{2}{7} \, \kappa_0 \, \frac{T_M^{2}}{n_M L} \,\,\, .
\end{equation}
It is important to remark here that the sound speed has been defined in terms of the maximum (apex) temperature and so the Mach number, in this context, is the Mach number which gives the 0D bulk flow speed with respect to the apex sound speed. Though temperature increases monotonically towards the apex, as does the sound speed ($C_S \propto \sqrt{T}$), the analysis of Section~\ref{validity} shows that the bulk flow cannot exceed $\simeq 77\%$ of the local sound speed anywhere along the loop. Writing

\begin{equation}\label{x-def}
x = \frac{T_M^2}{n_M L}
\end{equation}
gives a quadratic equation for $x$:

\begin{equation}\label{quadratic_for_x}
x^2 - K_1  \, \left ( \frac{M}{\gamma-1} + \frac{M^3}{2} \right ) x - K_2^2 = 0 \,\,\, ,
\end{equation}
where

\begin{equation}\label{k1-k2-def}
K_1 = \frac{7}{2 \kappa_0}\left ( \frac{8 \gamma^3 k_B^3}{m_p} \right )^{1/2}  \simeq 1.57 \times 10^{-5} \, ; \qquad K_2 = \left ( \frac{7 \, \chi}{2 \kappa_0} \right)^{1/2} \simeq 5.74 \times 10^{-7} \,\,\, ,
\end{equation}
both in the same units as $x$ (K$^2$~cm$^2$). We note that the ratio

\begin{equation}\label{k-ratio}
\frac{K_1}{K_2} \simeq 27 \,\,\, .
\end{equation}
The three terms on the left-hand side of Equation~(\ref{quadratic_for_x}) arise from thermal conduction, enthalpy, and radiation, respectively, and its physical (positive-root) solution is

\begin{equation}\label{quad-sol}
x = \frac{K_1}{2}  \, \left ( \frac{M}{\gamma-1} + \frac{M^3}{2} \right ) + \sqrt{\frac{K_1^2}{4}  \, \left ( \frac{M}{\gamma-1} + \frac{M^3}{2} \right )^2 + K_2^2} \,\,\, .
\end{equation}
Since Equation~(\ref{quad-sol}) determines the value of $x$ for a given Mach number $M$, it follows that {\it the scaling $T_M \propto (n_ML)^{1/2}$ is valid for any $M$}, including the static case $M=0$, when the radiative losses scale as $T^{-1/2}$.

\begin{figure}[pht]
	\centering
	\includegraphics[width=0.6\linewidth]{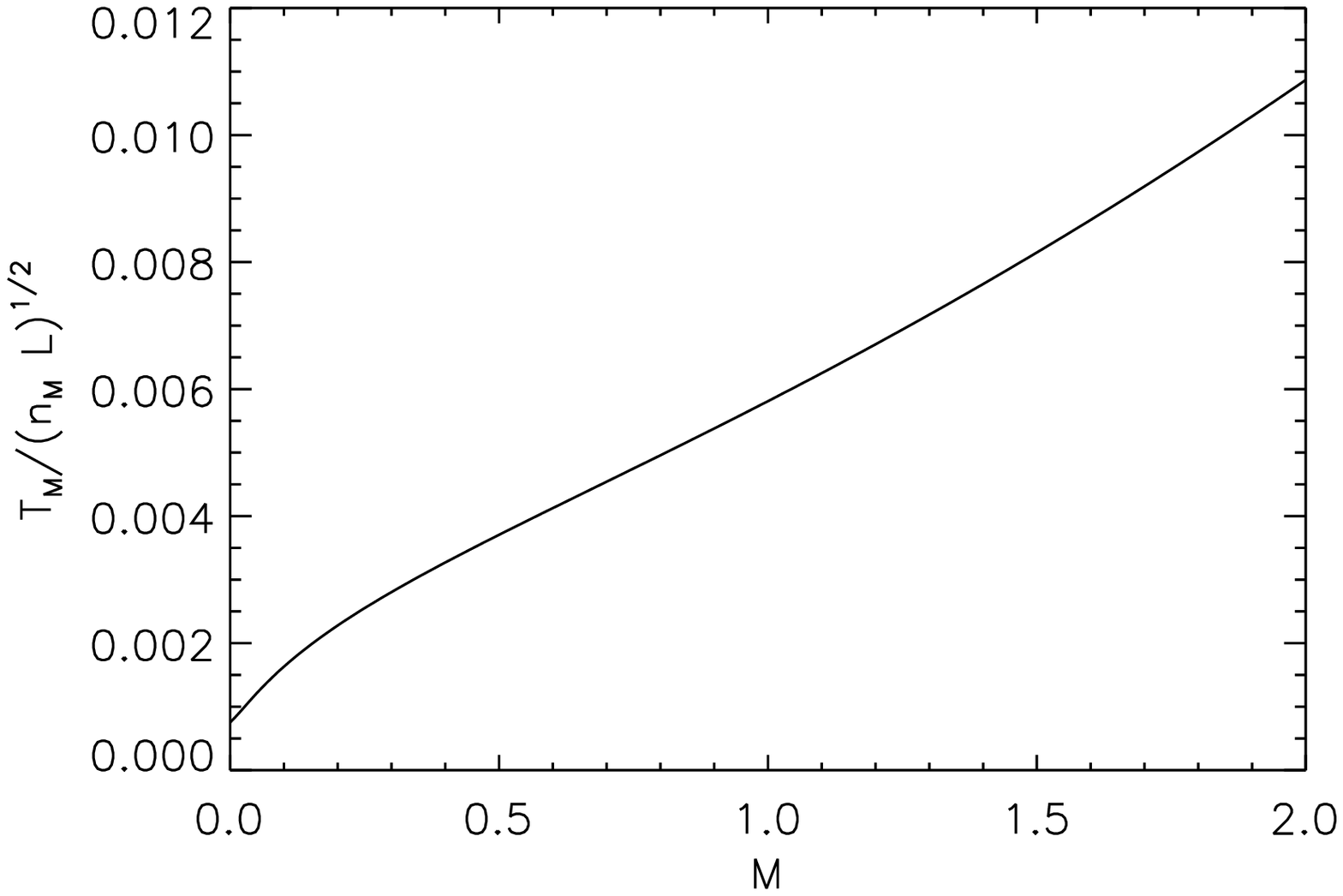}
	\caption{\label{fig:m-dep} Variation of the scaling law coefficient $T_M/(n_ML)^{1/2}$ with Mach number $M$ for heated loops (Equations~(\ref{x-def}) and~(\ref{quad-sol})).}
\end{figure}

The solid line in Figure~\ref{fig:m-dep} shows the scaling ratio $T_M/(n_ML)^{1/2}$ vs. Mach number $M$, and it is instructive to consider analytic approximations to this behavior in several regimes.

\begin{enumerate}
	
	\item For very slow flows
	
	\begin{equation}\label{very-slow-condition}
	0 \le M \lapprox \frac{2 (\gamma-1)}{(K_1/K_2)} \simeq 0.05 \,\,\, ,
	\end{equation}
	the dominant energy balance is between thermal conduction, radiation, and the thermal energy ($\propto M/(\gamma-1)$) component of the enthalpy flux, and the $K_2^2$ term in the radical in Equation~(\ref{quad-sol}) dominates. We then obtain
	
	\begin{equation}\label{x-sol-slow-flows}
	x \simeq K_2 \left ( 1 + \frac{(K_1/K_2)}{2(\gamma-1)} M \right ) \,\,\, ,
	\end{equation}
	which, using Equation~(\ref{x-def}), gives
	
	\begin{equation}\label{t-sol-very-slow}
	T_M \simeq K_2^{1/2} \left ( 1 + \frac{(K_1/K_2)}{2(\gamma-1)} M \right )^{1/2} \, (n_M L)^{1/2} \simeq 7.6 \times 10^{-4} \left ( 1 + 20 M \right )^{1/2} \, (n_M L)^{1/2} \,\,\, ,
	\end{equation}
	where $M=0$ in the static case.
	
	\item For subsonic flows

	\begin{equation}\label{slow-fast-transition}
	\frac{2 (\gamma - 1) }{(K_1/K_2)} \lapprox M \lapprox  \sqrt{2/(\gamma-1)} \, ; \,\,\, {\rm i.e.,} \,\,\, 0.05 \lapprox M \lapprox 0.77 \,\,\, ,
	\end{equation}
	 the dominant energy balance is between thermal conduction and the thermal energy component of the enthalpy flux, and the term proportional to $K_1$ in the radical in Equation~(\ref{quad-sol}) dominates.  However, we can still neglect the kinetic energy ($\propto M^3/2$) term, giving

	\begin{equation}\label{quad-sol-slow}
	x = K_1  \, \left ( \frac{M}{\gamma-1} \right ) \,\,\, ,
	\end{equation}
	so that, again using Equation~(\ref{x-def}),
	
	\begin{equation}\label{t-sol-low-M}
	T_M \simeq \left ( \frac{K_1}{\gamma - 1}  \right )^{1/2} \, M^{1/2} \, (n_M L)^{1/2} \simeq 4.9 \times 10^{-3} \, M^{1/2} \, (n_M L)^{1/2} \,\,\, .
	\end{equation}
\end{enumerate}

We reiterate that in all cases we obtain the same functional dependence $T_M \propto (n_ML)^{1/2}$ in the scaling law, but with different coefficients:

\begin{equation}\label{t-scalings}
T_M \simeq
\begin{cases}
7.6 \times 10^{-4} \, (n_M L)^{1/2} \qquad &; \qquad {\rm static}; M=0 \cr
7.6 \times 10^{-4} \left ( 1 + 20 M \right )^{1/2} \, (n_M L)^{1/2} \qquad &; \qquad 0 < M \lapprox 0.05 \cr
4.9 \times 10^{-3} \, M^{1/2} \, (n_M L)^{1/2} \qquad &; \qquad 0.05 \lapprox M \lapprox 0.77 \cr
\end{cases}
\,\,\, .
\end{equation}
Further, {\it for low Mach number flows $M \lapprox 0.05$, the scaling law is almost indistinguishable from the result for a static loop.} In the supersonic case $M \sim 1$ the dominant energy balance would be between thermal conduction and the kinetic energy ($\propto M^3/2$) component of the enthalpy flux, according to the complete scaling law (\ref{quad-sol}), though of course we do not consider this limit for the reasons discussed in Section~\ref{validity}.

We wish to stress that this remarkable result, that the scaling $T_M \propto (n_ML)^{1/2}$ applies to both static and dynamic loops, is not obvious {\it a priori}. For example, if the radiative loss term $\Lambda(T)$ had a different temperature dependence (e.g., $\Lambda(T) = \chi_* T^{-\delta}; \delta \ne 1/2$) in the region of interest, then Equation~(\ref{energy-equation-red}) would have taken on the form

\begin{equation}\label{energy-equation-red-delta}
\left ( \frac{8 \gamma^3 k_B^3}{m_p} \right )^{1/2}  \, \left ( \frac{M}{\gamma-1} + \frac{M^3}{2} \right ) + \chi_* \, n_M \, T_M^{-3/2 - \delta} \, L  = \frac{2}{7} \, \kappa_0 \, \frac{T_M^{2}}{n_ML} \,\,\, .
\end{equation}
Making the substitution

\begin{equation}\label{x-general}
x = \frac{T^{\frac{7}{4} + \frac{\delta}{2} }}{n_ML}
\end{equation}
and setting

\begin{equation}\label{k1-k2-def-general}
K_1 = \frac{7}{2 \kappa_0}\left ( \frac{8 \gamma^3 k_B^3}{m_p} \right )^{1/2}  ({\rm as \, before})  \, ; \qquad K_2 = \left ( \frac{7 \, \chi_*}{2 \kappa_0} \right )^{1/2} \,\,\, ,
\end{equation}
gives

\begin{equation}\label{quadratic}
x^2 - K_1  \, \left ( \frac{M}{\gamma-1} + \frac{M^3}{2} \right ) T_M^{(\delta - 1/2)/2} \, x - K_2^2 = 0 \,\,\, .
\end{equation}
The appearance of $T^{(\delta - 1/2)/2}$ in the second term on the left-hand side of Equation~(\ref{quadratic}) means this is no longer a simple quadratic equation in $x$ when $\delta \ne 1/2$. {\it Only for the static case} does this troublesome term vanish, leading to a scaling law $x = K_2$ or

\begin{equation}\label{static-scaling-general-delta}
T_M \propto (n_ML)^{\frac{4}{7+2 \delta}} \,\,\, ;
\end{equation}
however, a similarly straightforward scaling law cannot be found for the dynamic case $M \ne 0$.

\subsection{Heating, Density, and Length}
\label{scaling_law_2}

The second scaling law in \cite{1978ApJ...220..643R}, given by Equation~(\ref{rtv44}) above, is found by equating the volumetric heating and the radiative losses at the loop apex, such that the heat flux at the upper boundary is set to zero, and substituting the first scaling law for the temperature. In the static case, the energy input to the corona is redistributed by thermal conduction into the lower chromospheric layers of the atmosphere, where it is radiated away.  By contrast, in the dynamic case this downward flow of energy is more than can be radiated away, which results in a return of energy to the corona via an upward enthalpy flux. The energy balance in the dynamic case is thus given by

\begin{equation}\label{energy_equation_2}
\frac{dF_E}{ds} + E_R = E_H \,\,\, ,
\end{equation}
where $E_H$ is the (assumed uniform) volumetric heating rate. We see that the left-hand side of Equation~(\ref{energy_equation_2}) is the same as Equation~(\ref{energy_equation}) so we can simply write

\begin{equation}\label{energy_equation_3}
E_H = -\frac{dF_C}{ds} = \frac{2}{7} \, \kappa_0 \, \frac{T_M^{7/2}}{L^2} \,\,\, ,
\end{equation}
and substituting for $T_M$ using Equation~(\ref{x-def}) gives

\begin{equation}\label{energy_equation_4}
E_H = \frac{2}{7} \, \kappa_0 \, x^{7/4} \, n_M^{7/4} \, L^{-{1/4}} \,\,\, ,
\end{equation}
where $x$ is given by Equation~(\ref{quad-sol}).

We consider the same limiting cases as in Section~\ref{scaling_law_1}.

\begin{enumerate}

\item For very slow flows

\begin{equation}\label{h-sol-very-slow}
E_H \simeq \frac{2}{7} \, \kappa_0 \, K_2^{7/4} \left ( 1 + \frac{(K_1/K_2)}{2(\gamma-1)} M \right )^{7/4} \, n_M^{7/4} \, L^{-1/4} \simeq 5.8 \times 10^{-18} \left ( 1 + 20 M \right )^{7/4} \, n_M^{7/4} \, L^{-1/4} \,\,\, ,
\end{equation}
where $M=0$ in the static case.

\item For subsonic flows

\begin{equation}\label{slow-fast-transition2}
E_H \simeq \frac{2}{7} \, \kappa_0 \left( \frac{K_1}{\gamma-1} \right)^{7/4} M^{7/4} \, n_M^{7/4} \, L^{-1/4} \simeq 2.4 \times 10^{-15} M^{7/4} \, n_M^{7/4} \, L^{-1/4} \,\,\, .
\end{equation}

\end{enumerate}
Once again, we note that in all cases we obtain the same functional dependence $E_H \propto n_M^{7/4} L^{-1/4}$, but with different coefficients:

\begin{equation}\label{}
E_H \simeq
\begin{cases}
5.8 \times 10^{-18} \, n_M^{7/4} L^{-1/4} \qquad &; \qquad {\rm static}; M=0 \cr
5.8 \times 10^{-18} \left ( 1 + 20 M \right )^{7/4} \, n_M^{7/4} \, L^{-1/4} \qquad &; \qquad 0 < M \lapprox 0.05 \cr
2.4 \times 10^{-15} \, M^{7/4} \, n_M^{7/4} \, L^{-1/4} \qquad &; \qquad 0.05 \lapprox M \lapprox 0.77 \cr
\end{cases}
\,\,\, .
\end{equation}
The energy input ($E_H$) increases monotonically with the Mach number since stronger heating drives a larger downward conductive flux and consequently a stronger return enthalpy flux into the corona.

\section{Turbulence-Dominated Conduction}\label{turb-dom}

During the dynamic process of coronal heating it is highly likely that some form of turbulence exists in the loop plasma, which can fundamentally change the form of the heat flux term in the energy equation \citep{2018ApJ...852..127B}.  \cite{2019ApJ...880...80B} have shown that this can dramatically alter the form of the scaling laws appropriate to quasi-static loops. Here we explore the effects of turbulence-dominated thermal conduction on the scaling laws appropriate to dynamic loops.

For transport dominated by turbulence, the appropriate form for $\lambda$ is the turbulent mean free path $\lambda_T$ which, following \cite{2018ApJ...852..127B}, we take as independent of temperature.  For such a case, Equation~(\ref{kappa-general}) gives $\kappa \propto n \, T^{1/2}$, so that the conduction term~(\ref{cond-general}) becomes

\begin{equation}\label{conduction-turbulent}
- \frac{2 k_B \kappa_0 \, \lambda_T \, n}{c_R}  \, T^{1/2} \, \frac{dT}{ds} \simeq - \frac{2 k_B \kappa_0 \, \lambda_T \, n_M}{c_R}  \, \frac{T_M^{3/2}}{L} \,\,\, ,
\end{equation}
where

\begin{equation}\label{cr-def}
c_R = \frac{4 k_B^3}{\pi e^4 \ln \Lambda}
\end{equation}
\citep{2019ApJ...880...80B}.  At first sight, it appears that we could simply use this amended expression for the heat flux in the energy equation~(\ref{energy_equation}) and integrate over the loop half-length $L$. Such an (unfortunately naive; see below) approach results in the zero-dimensional energy equation (cf. Equation~(\ref{0-D}))

\begin{equation}\label{}
\left ( \frac{8 \gamma^3 k_B^3}{m_p} \right )^{1/2}  \, \left ( \frac{M}{\gamma-1} + \frac{M^3}{2} \right ) n_M \, T_M^{3/2} + \chi \, n_M^2 \, T_M^{-1/2} \, L  = \frac{2 k_B \kappa_0 \, \lambda_T \, n_M}{c_R}  \, \frac{T_M^{3/2}}{L} \,\,\,
\end{equation}
or

\begin{equation}\label{}
\left ( \frac{8 \gamma^3 k_B^3}{m_p} \right )^{1/2}  \, \left ( \frac{M}{\gamma-1} + \frac{M^3}{2} \right ) + \chi \, n_M \, T_M^{-2} \, L  = \frac{2 k_B \kappa_0 \, \lambda_T}{c_R L} \,\,\, .
\end{equation}
The scaling law thus depends on the value of the (small, but finite) turbulent Knudsen number ${\rm Kn_T} = \lambda_T/L$. With $x=T_M^2/n_ML$ as before, we find

\begin{equation}\label{x-turb}
x = \frac{\chi}{ \frac{2 k_B \kappa_0 \, \lambda_T }{c_R \, L} - \left ( \frac{8 \gamma^3 k_B^3}{m_p} \right )^{1/2}  \, \left ( \frac{M}{\gamma-1} + \frac{M^3}{2} \right ) }\,\,\, .
\end{equation}

For the static case, setting $M=0$ gives

\begin{equation}\label{x-static-turb}
x = \frac{\chi \, c_R \, L}{ 2 k_B \, \kappa_0 \, \lambda_T }\,\,\,
\end{equation}
resulting in the (tentative) scaling law

\begin{equation}\label{turb-scale-tentative}
T_M^2 = \frac{\chi \, c_R }{ 2 k_B \, \kappa_0 \, \lambda_T }  \, n_M L^2 \, ; \qquad T_M = \left ( \frac{\chi \, c_R }{ 2 k_B \, \kappa_0 \, \lambda_T } \right )^{1/2} \, (n_M L^2)^{1/2} \,\,\, .
\end{equation}
However, \cite{2019ApJ...880...80B} have shown that in the static case, the much weaker scaling of the thermal conduction coefficient with temperature (for a given $n$, $\kappa \propto T^{1/2}$, while for a given $P$, $\kappa \propto T^{-1/2}$) leads to a dominant role for the {\it base} (rather than peak) temperature and consequently also in the resulting scaling law. Taking this into account, they formally solved the energy equation and found a rigorous solution for the static scaling law (Equation~(23) in that paper)

\[T_M = \frac{\chi c_R}{64 k_B^2 \, \kappa_0  \lambda_T \, T_{0}^2} \, P L^2 \,\,\, ,\]
which, since we are working in terms of $n_M$ rather than $P$, we rewrite as


\begin{equation}\label{equation-23}
n_M = \frac{ 32 k_B \, \kappa_0}{ \chi \, c_R } \, \frac{\lambda_T}{L} \, \frac{T_0^2}{L} \,\,\, .
\end{equation}
Equation~(\ref{equation-23}) differs significantly from the scaling law~(\ref{turb-scale-tentative}) that was obtained by naively balancing thermal conduction, radiation, and enthalpy at a single temperature. The major difference is that the quantity $T_M$ no longer appears in the scaling law relating density $n_M$ and loop half-length $L$, due to the proportionality between $T_M$ and $P$ in the original scaling law (see Equation~(\ref{T_M_P_M_L}) in Appendix~\ref{app_b}). We now have a situation where the electron number density $n_M$ at the loop apex scales with the turbulent Knudsen number ${\rm Kn_T} = \lambda_T/L$ and has a strong, non-linear, dependence on the base temperature. There is also the expected inverse relationship with the loop length, where longer loops have lower apex densities.

We can transform Equation~(\ref{turb-scale-tentative}) into a form consistent with Equation~(\ref{equation-23}) by defining $T_* = 16T_0^2/T_M$ and replacing $T_M^2$ on the left-hand side of Equation~(\ref{turb-scale-tentative}) with $T_M T_* = 16 T_0^2$; this yields the correct static scaling law. We therefore, by analogy, replace $T_M^2$ with $16 T_0^2$ in the definition for $x$:

\begin{equation}\label{gen-n-turb}
x = 16 \frac{T_0^2}{n_M L} \,\,\, .
\end{equation}

The general scaling law for the uniform volumetric heating rate can be found by replacing the collisionally-dominated heat flux~(\ref{conduction-spitzer}) with the turbulence-dominated form~(\ref{conduction-turbulent}) in the energy equation~(\ref{energy_equation_3}):

\begin{equation}
E_H = \frac{2 k_B \kappa_0 \lambda_T n_M}{c_R} \, \frac{T_M^{3/2}}{L^2} \,\,\, .
\end{equation}
Substituting for $n_M$ from Equation~(\ref{gen-n-turb}) gives

\begin{equation}\label{gen-h-turb}
E_H = \frac{32 k_B \kappa_0}{c_R} \, \frac{\lambda_T}{L} \, \frac{T_0^2 \, T_M^{3/2}}{x L^2} \,\,\, .
\end{equation}
In the static case, using Equation~(\ref{x-static-turb}) for $x$, we obtain

\begin{equation}\label{static-h-turb}
E_H = \frac{1}{\chi} \, \left( \frac{8 k_B \kappa_0}{c_R} \right)^2 \, \left( \frac{\lambda_T}{L} \right)^2 \frac{T_0^2 \, T_M^{3/2}}{L^2} = \frac{128 k_B^3}{m_e} \, \frac{1}{\chi} \,  \left( \frac{\lambda_T}{L} \right)^2 \frac{T_0^2 \, T_M^{3/2}}{L^2} \,\,\, ,
\end{equation}
where we have used Equations~(\ref{kappa-spitzer}) and~(\ref{cr-def}) for the ratio $\kappa_0/c_R$.

In summary, the maximum temperature $T_M$ has no apparent dependence on either the apex number density $n_M$ and the loop half-length $L$ (Equation~(\ref{equation-23})).  For a given turbulent mean free path $\lambda_T$, the apex number density scales as the square of the quantity $T_0/L$.  The volumetric heating rate $E_H$ depends on the loop half-length and a combination of the base and maximum temperatures.

What is the physical explanation for these scaling laws? First, to understand why $T_M$ and $n_M$ are seemingly independent, consider the extreme case of very strong electron scattering by turbulence, corresponding to very small values of ${\rm Kn_T}$, which effectively prevents any heat flux from escaping the corona. In this situation, regardless of whether $T_M$ continued to increase, the heat flux would remain trapped and consequently, with no conduction-driven ablation of chromospheric material, there would be no density change. $n_M$ thus depends only on the pre-existing conditions in the loop, as given by Equation~(\ref{gen-n-turb}).

On the other hand, the scaling law for $E_H$ {\it does} depend upon $T_M$. This is because $E_H$ is found by balancing heating, radiation, and the divergence of the enthalpy flux at the temperature maximum, which is equivalent to balancing heating and thermal conduction: cf. Equations~(\ref{energy_equation}) and~(\ref{energy_equation_2}). Since the radiative loss function is proportional to the square of the apex density $n_M$ and also depends on the temperature $T_M$, so does the required volumetric heating rate. Since the density $n_M$ in turn depends on both $T_0$ and $L$, those quantities also appear in the scaling law~(\ref{static-h-turb}). Further, because the apex density $n_M$ is inversely proportional to the radiative loss coefficient $\chi$ (Equation~(\ref{equation-23})), the total radiative losses per unit volume and, consequently, the volumetric heating rate, scale as $(1/\chi)^2 \, \chi = 1/\chi$, as in Equation~(\ref{static-h-turb}). The relationship $E_H \propto 1/\chi$ arises because a larger value of $\chi$ increases the radiative losses, removing more energy from the incoming heat flux and leaving less for the return enthalpy flux, which in turn yields a lower coronal density. The reduced radiative losses at the loop apex, where the scaling law for $E_H$ is derived, are then sustained by a lower volumetric heating rate.

We note linear and quadratic dependencies on the turbulent Knudsen number ${\rm Kn_T}$ for $n_M$ and $E_H$, respectively. As ${\rm Kn_T}$ decreases, and the flux-limiting effect strengthens, the heat flux becomes increasingly less able to fill the loop via ablation from the lower atmosphere. In addition, less energy input is required to sustain the corona against its radiative losses, due to the increasing amount of energy becoming bottled-up by the strong turbulence and unable to escape.

We also notice that the expression~(\ref{x-turb}) becomes infinite at a value of the Mach number $M$ given by

\begin{equation}\label{m-max}
\left( \frac{M}{\gamma-1} + \frac{M^3}{2} \right) = \frac{1}{\gamma^{3/2}} \left( \frac{m_p}{m_e} \right)^{1/2} \frac{\lambda_T}{L} \,\,\, ,
\end{equation}
so that $M$ is bounded above by a limit which decreases as the turbulent Knudsen number decreases, i.e., the heat flux is suppressed to a greater extent. At the upper limit of $M$, where $x \rightarrow \infty$ (Equation~(\ref{x-turb})), Equations~(\ref{gen-n-turb}) and (\ref{gen-h-turb}) would both imply that the number density and volumetric heating rate go to zero, which is of course not physical.

For small (but not {\it very} small) values of the turbulence scale length $\lambda_T$ (weakly suppressed conduction), the cubic $M^3$ term on the left-hand side of Equation~(\ref{m-max}) dominates, giving a limiting Mach number

\begin{equation}\label{limiting-M-large}
M_{\rm max} \simeq \frac{2^{1/3}}{\gamma^{1/2}} \left( \frac{m_p}{m_e} \right)^{1/6} \, \left ( \frac{\lambda_T}{L} \right )^{1/3} \simeq 3.4 \, \left( \frac{\lambda_T}{L} \right)^{1/3} \,\,\, .
\end{equation}
For even smaller values of $\lambda_T/L$ the linear $M$ term on the left-hand side dominates, giving a limiting Mach number

\begin{equation}\label{limiting-M-small}
M_{\rm max} \simeq \frac{\gamma-1}{\gamma^{3/2}} \left( \frac{m_p}{m_e} \right)^{1/2} \, \frac{\lambda_T}{L} \simeq 13.3 \, \frac{\lambda_T}{L} \,\,\, .
\end{equation}

Expressions~(\ref{limiting-M-large}) and~(\ref{limiting-M-small}) are equal when Kn$_T \simeq 0.13$. Thus,

\begin{itemize}
	
	\item when the turbulent conduction limitation is finite but not too severe ($0.13 \lapprox {\rm Kn}_T \lapprox 1$), the cube-root scaling~(\ref{limiting-M-large}) applies;
	
	\item in the strongly turbulence-limited case ${\rm Kn}_T \lapprox 0.13$, the linear scaling~(\ref{limiting-M-small}) applies.
	
\end{itemize}

\begin{figure}[pht]
	\centering
	\includegraphics[width=0.6\linewidth]{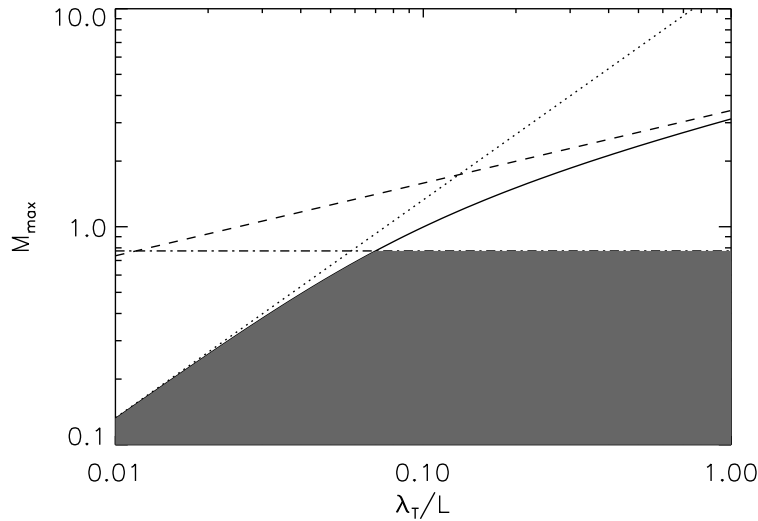}
	\caption{Variation of the maximum Mach number $M$ with the turbulent Knudsen number ${\rm Kn_T}=\lambda_T/L$. Equation~(\ref{m-max}) is shown by the solid line, and the limiting cases for large ${\rm Kn_T}$ (Equation~(\ref{limiting-M-large})) and small ${\rm Kn_T}$ (Equation~(\ref{limiting-M-small})) are shown by the dashed and dotted lines, respectively. The horizontal dashed-dot line indicates the limit $M=\sqrt{1/\gamma} \simeq 0.77$ imposed by hydrodynamic considerations (Equation~(\ref{mmax-hydro})), which essentially restricts flows to the linear regime of Equation~(\ref{limiting-M-small}). Thus, below the corresponding value of the turbulent Knudsen number ${\rm Kn}_T \simeq 0.06$ (Equation~(\ref{knt-threshold})), the flow speed is limited by turbulent suppression of the driving conductive flux to the $M \propto {\rm Kn}_T$ regime of Equation~(\ref{limiting-M-small}); above ${\rm Kn}_T \simeq 0.06$, it is limited by hydrodynamic considerations to $M=\sqrt{1/\gamma}$. The shaded area shows the range of allowed Mach number values over the range of turbulent Knudsen numbers ${\rm Kn}_T$. }\label{fig:msolve}
\end{figure}

Figure~\ref{fig:msolve} shows the variation of $M_{\rm max}$ with ${\rm Kn_T}$, together with the linear and cube-root limits discussed above. The transition between regimes occurs as $M_{\rm max}$ decreases below $\simeq 1.7$. However we have noted in Section~\ref{validity} that the bulk flow velocity is hydrodynamically limited to $M_{\rm max} =\sqrt{1/\gamma} \simeq 0.77$.  Thus only in the strongly limited case

\begin{equation}\label{knt-threshold}
\frac{\lambda_T}{L} < \frac{1}{13.3\sqrt{\gamma}} \simeq 0.06
\end{equation}
is the maximum Mach number $M_{\rm max}$ of the bulk flow set by (and indeed roughly proportional to) the turbulent Knudsen number ${\rm Kn_T}$; for higher values of ${\rm Kn_T}$, $M_{\rm max}$ is established by the hydrodynamic limit $\sqrt{1/\gamma}$.

When thermal conduction $\propto \kappa \, dT/ds$ is dominated by collisions, the conduction coefficient $\kappa$ scales as a fairly large positive power ($T^{5/2}$) of the temperature $T$, so that loops with larger apex temperatures $T_M$ drive considerably more heat flux into the transition region. Now, because of the high velocity ratio $\sqrt{m_p/m_e}$ between electron-driven thermal fronts and ion-driven, information-carrying, sound waves driven by the pressure changes associated with apex temperature changes, the transition region is heated before being compressed by the increased coronal pressure. Since the radiative losses $E_R$ are proportional to the square of the density, the still tenuous transition region is unable to radiate away the incoming heat flux when it arrives and the local pressure increase drives a return enthalpy flux back into the corona. Thus, in a collision-dominated regime, even a modest increase in the loop apex temperature $T_M$ produces a very significant increase in the conductive flux, and increases the Mach number of the flow that returns excess energy to the corona. The maximum loop temperature $T_M$ straightforwardly correlates positively with the Mach number in every case (Equations~(\ref{t-scalings})).

With this collisional context established, the physical reason for the reduced $M_{\rm max}$ in the presence of turbulence is clear. An increase in the Mach number of the return enthalpy flux requires an increased heat flux $\kappa \, dT/ds$ to drive it. But in the turbulent regime $\kappa$ has a much weaker dependence on temperature ($\propto T^{1/2}$) and furthermore the temperature gradient is limited by the magnitude of the loop half-length $L$ over which the corona-chromosphere temperature difference exists. It follows that the downward heat flux, and hence the speed of the return enthalpy flow, is necessarily limited. Furthermore, the lower the turbulent Knudsen number ${\rm Kn_T}$, the higher the suppression of the heat flux, and hence the lower the Mach number of the return enthalpy flow; accordingly $M_{\rm max}$ scales with ${\rm Kn_T}$ (Equation~(\ref{limiting-M-small})).

These considerations establish a natural upper limit on the Mach number of the return flow that can be driven by turbulence-limited thermal conduction, established by the physical scale of the turbulent interactions.

\section{Summary and Conclusions}\label{conclusions}

We have extended the static loop scaling-law analysis of \cite{1978ApJ...220..643R} to the case of dynamic loops in which the upward enthalpy flux associated with excess transition region energy, that cannot be efficiently radiated, plays an important role in the energy equation. We have also constrained the range of validity of these dynamic scaling laws to Mach numbers $0 \le M \le \sqrt{1/\gamma}$, where the upper limit corresponds to a flow at the ion thermal speed. Quite remarkably, we find, in the case where thermal conduction is characterized by a mean free path $\lambda$ associated with Coulomb collisions \citep{1962pfig.book.....S}, that {\it the functional forms of the loop scaling laws, viz. $T_M \propto (n_ML)^{1/2}$ and $E_H \propto n_M^{7/4} L^{-1/4}$ ($T_M \propto (P_ML)^{1/3}$ {\rm and} $E_H \propto P_M^{7/6}L^{-5/6}${\rm; see Appendix~\ref{app_a})}, are maintained}. We have further stressed that this result, far from being obvious {\it a priori}, is due to the serendipitous temperature dependence of the radiative loss term in the domain of interest. Further, for low Mach number flows, the constant of proportionality ($\propto ( 1 + 20 M )^{1/2}$) differs only very mildly from the static case (Equation~(\ref{t-scalings}) and Figure~\ref{fig:m-dep}), further extending the domain of applicability of the static \cite{1978ApJ...220..643R} scaling law.

For heat transport that is controlled by a turbulent scale length $\lambda_T$ independent of temperature, there is a three-powers-of $T$ difference between the conduction coefficient and that for the collision-dominated case $\kappa$ ($\lambda_T \sim$ constant vs. $\lambda_C \propto T^2/n \propto T^3/P$).  This changes the temperature dependence of the coefficient $\kappa$ in the thermal conduction term $\kappa \, dT/ds$ from a strong positive dependence on temperature ($\kappa \propto T^{5/2}$) to a weak positive dependence ($\kappa \propto T^{1/2}$) for constant density and a weak {\it negative} dependence ($\kappa \propto T^{-1/2}$) for constant pressure $P$. Because of this, larger conductive fluxes no longer straightforwardly result from a modest increase in peak loop temperature $T_M$; instead they require a very significant increase in the local temperature gradient.  Accordingly, a putative high Mach-number upflow requires a temperature gradient that is incompatible with the hierarchy of scales demanded for turbulence to be an effective heat flux limiting mechanism, and hence such flows cannot occur. The resulting upper limit on the Mach number associated with the upward flow of enthalpy (Figure~\ref{fig:msolve}) depends on the extent to which thermal conduction is suppressed, i.e., on the turbulent Knudsen number ${\rm Kn_T} = \lambda_T/L$.  For very small turbulent Knudsen numbers ${\rm Kn_T}$, the limiting Mach number is proportional to ${\rm Kn_T}$ (Equation~(\ref{limiting-M-small})), while for higher turbulent Knudsen numbers, the limiting Mach number is determined by hydrodynamic considerations (Equation~(\ref{mmax-hydro})).

\appendix

\section{Dynamic Scaling Laws Cast in Terms of Pressure and Length}
\label{app_a}

The scaling laws have traditionally been written with respect to pressure and length \citep[e.g.,][]{1978ApJ...220..643R,2010ApJ...714.1290M,2019ApJ...880...80B} on the basis that in the static case the pressure can be taken as constant in the field-aligned direction (neglecting gravity). This makes pressure a convenient variable, but the assumption does not hold in the dynamic case for the reasons discussed in Section~\ref{validity}, and so we have elected to pursue the derivation of dynamic scaling laws with respect to number density and length.

\begin{figure}[pht]
	\centering
	\includegraphics[width=0.6\linewidth]{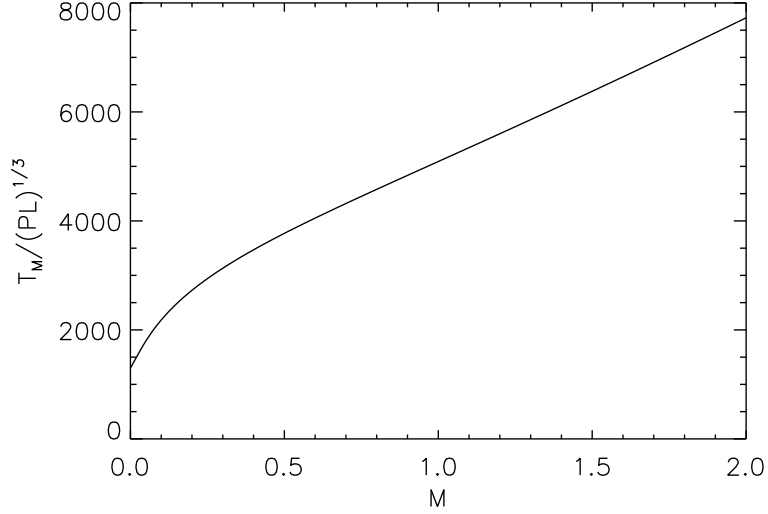}
	\caption{\label{fig:m-dep-P} Variation of the scaling law coefficient $T_M/(PL)^{1/3}$ with Mach number $M$ for heated loops (Equation~(\ref{TM_P_scaling})).}
\end{figure}

Nonetheless, we provide the corresponding formulations with respect to pressure and length below for collision-dominated thermal conduction, and in Appendix~\ref{app_b} for turbulence-dominated thermal conduction. We emphasize that pressure is now the value at the maximum (apex) temperature $P_M \equiv P(T_M)$.

Equation~(\ref{quad-sol}) gives the complete solution for the scaling law for $T_M$, as before, where the variables $x$, $K_1$ and $K_2$ become

\begin{equation}\label{x_def_P}
x = \frac{T_M^3}{P_ML} \,\,\, ; \qquad K_1 = \frac{7 }{2 \kappa_0}\left ( \frac{2 \gamma^3 k_B}{m_p} \right )^{1/2}  \simeq 6 \times 10^{10} \,\,\, ; \qquad K_2 = \frac{1}{2 k_B} \, \sqrt{\frac{7 \, \chi}{2 \kappa_0} } \simeq 2 \times 10^9 \,\,\, .
\end{equation}
The scalings of $T_M$ and $E_H$ with respect to $P_M$ and $L$ are all the same as in the static cases, with the dynamic modifications entering via Mach number-dependent coefficients.

\begin{equation}\label{TM_P_scaling}
T_M \simeq
\begin{cases}
1300 \, (P_ML)^{1/3} \qquad &; \qquad {\rm static}; M=0 \cr
1300 \left ( 1 + 20 M \right )^{1/3} \, (P_ML)^{1/3} \qquad &; \qquad 0 < M \lapprox 0.05 \cr
4500 \, M^{1/3} \, (P_ML)^{1/3} \qquad &; \qquad 0.05 \lapprox M \lapprox 0.77
\end{cases}
\,\,\, .
\end{equation}

\begin{equation}
E_H \simeq
\begin{cases}
3.45 \times 10^4 \, P_M^{7/6} L^{-5/6} \qquad &; \qquad {\rm static}; M=0 \cr
3.45 \times 10^4 \left ( 1 + 23 M \right ) \, P_M^{7/6} \, L^{-5/6} \qquad &; \qquad 0 < M \lapprox 0.05 \cr
2.93 \times 10^6 \, M^{7/6} \, P_M^{7/6} \, L^{-5/6} \qquad &; \qquad 0.05 \lapprox M \lapprox 0.77
\end{cases}
\,\,\, .
\end{equation}
There are very small differences in the coefficients for the static cases between our formulae and RTV due to the slightly different value of $\ln \Lambda$ and consequently $\kappa_0$ that we have used.

\section{Dynamic Scaling Laws Cast in Terms of Pressure and Length for Turbulence-Dominated Conduction}
\label{app_b}

The complete solution for the scaling law for $T_M$ with respect to pressure and length in the turbulent case is given by (cf. Equation~(\ref{x-turb}))

\begin{equation}\label{x-app}
x = \frac{\left ( \frac{1}{2 k_B} \right )^2 \chi}{ \frac{2 \kappa_0 \, \lambda_T }{c_R \, L} - \left ( \frac{2 \gamma^3 k_B}{m_p} \right )^{1/2}  \, \left ( \frac{M}{\gamma-1} + \frac{M^3}{2} \right ) }\,\,\, ,
\end{equation}
which by direct (naive) substitution for $x$ results in the static scaling law

\begin{equation}\label{new-turb-law}
T_M^3 = \frac{ \chi \, c_R }{ 8 k_B^2 \, \kappa_0 \, \lambda_T }  \, P_M L^2 \, ; \qquad T_M = \left ( \frac{ \chi \, c_R }{ 8 k_B^2 \, \kappa_0 \, \lambda_T } \right )^{1/3} \, (P_M L^2)^{1/3} \,\,\, .
\end{equation}
Clearly, this is not consistent with Equation~(\ref{equation-23}).
By replacing $T_M^3$ with $8 T_M T_0^2$ in $x$, we recover the correct scaling law when this is substituted into Equation~(\ref{new-turb-law}). Thus

\begin{equation}\label{T_M_P_M_L}
T_M = \frac{ \chi \, c_R }{64 k_B^2 \, \kappa_0 \, T_0^2} \, \frac{L}{\lambda_T} \, P_M L \,\,\, .
\end{equation}
In general

\begin{equation}\label{}
T_M = \frac{x}{8T_0^2} P_M L \,\,\, .
\end{equation}

The general scaling law for the uniform volumetric heating rate is given by

\begin{equation}\label{eh-p-L}
E_H = \frac{\kappa_0 \lambda_T P_M}{c_R} \frac{T_M^{1/2}}{L^2} = \frac{\kappa_0}{c_R} \, \left(\frac{x}{8T_0^2}\right)^{1/2} \, \frac{\lambda_T}{L} \, P_M^{3/2} L^{-1/2} = \left(\frac{k_B \, x}{4 m_e \, T_0^2}\right)^{1/2} \, \frac{\lambda_T}{L} \, P_M^{3/2} L^{-1/2} \,\,\, .
\end{equation}

In the static case

\begin{equation}\label{}
E_H = \left( \frac{\chi \kappa_0}{64 c_R k_B^2 T_0^2} \right)^{1/2} \left( \frac{\lambda_T}{L} \right)^{1/2} P_M^{3/2} L^{-1/2} \,\,\, .
\end{equation}

\acknowledgements

SJB is grateful to the NSF for supporting this work through CAREER award AGS-1450230.  AGE was supported by grant NNX17AI16G from NASA's Heliophysics Supporting Research program. The authors would like to thank Professor Peter Cargill for extremely helpful discussions and suggestions during the preparation of this manuscript.

\bibliographystyle{aasjournal}
\bibliography{bib-scaling-laws}

\end{document}